\pdfoutput=1
\documentclass[11pt]{article}
\usepackage{authblk}

\usepackage{acl}

\usepackage{ulem}

\usepackage{times}
\usepackage{latexsym}

\usepackage[T1]{fontenc}
\usepackage[utf8]{inputenc}

\usepackage{microtype}

\usepackage{graphicx}
\usepackage{multirow}
\usepackage{subcaption}

\def\AAdel#1{\bgroup\markoverwith{\textcolor{blue}{\rule[0.5ex]{2pt}{1pt}}}\ULon{#1}}

\def\MSdel#1{\bgroup\markoverwith{\textcolor{red}{\rule[0.5ex]{2pt}{1pt}}}\ULon{#1}}

\title{Factors Affecting the Performance of Automated Speaker Verification in Alzheimer’s Disease Clinical Trials}

\author{\bf{Malikeh Ehghaghi, Marija Stanojevic, Ali Akram, Jekaterina Novikova}\\
        Winterlight Labs Inc., Toronto, Canada \\
        \texttt{\{malikeh, marija, aliakram, jekaterina\}@winterlightlabs.com}\\}

\begin{document}
\maketitle
\begin{abstract}
Detecting duplicate patient participation in clinical trials is a major challenge because repeated patients can undermine the credibility and accuracy of the trial's findings and result in significant health and financial risks. Developing accurate automated speaker verification (ASV) models is crucial to verify the identity of enrolled individuals and remove duplicates, but the size and quality of data influence ASV performance.
However, there has been limited investigation into the factors that can affect ASV capabilities in clinical environments. In this paper, we bridge the gap by conducting analysis of how participant demographic characteristics, audio quality criteria, and severity level of Alzheimer's disease (AD) impact the performance of ASV utilizing a dataset of speech recordings from 659 participants with varying levels of AD, obtained through multiple speech tasks. Our results indicate that ASV performance: 1) is slightly better on male speakers than on female speakers; 2) degrades for individuals who are above 70 years old; 3) is comparatively better for non-native English speakers than for native English speakers; 4) is negatively affected by clinician interference, noisy background, and unclear participant speech; 5) tends to decrease with an increase in the severity level of AD. Our study finds that voice biometrics raise fairness concerns as certain subgroups exhibit different ASV performances owing to their inherent voice characteristics. Moreover, the performance of ASV is influenced by the quality of speech recordings, which underscores the importance of improving the data collection settings in clinical trials.

\end{abstract}

\section{Introduction}
% Speech is widely studied as an automated biomarker for one's cognitive state \citep{fraser2016linguistic, boschi2017connected}, which presents a promising avenue toward automated assessment and monitoring of cognitive impairment in medical applications. With the rapid advancements in deep learning (DL), data-driven approaches have become a popular area of research in speech processing for detection of Alzheimer's disease (AD) and other types of dementia \citep{ebrahimighahnavieh2020deep}. 

% Accessing speech data for computer-assisted methods in healthcare can be challenging as voice biometrics can potentially identify speakers
% sharing recordings leading to patient privacy concerns. Because of the limited availability of data, many studies tend to focus on small cohorts \citep{maier2009speech}. 

% \sout{Conducting widespread longitudinal clinical trials, involving numerous patients, doctors, clinics, and even different countries can pose significant challenges in identifying instances of duplicate participation, which may ultimately jeopardize the validity of study findings. \citep{shiovitz2013cns} discovered that as much as 7.78\% of patients involved in large clinical trials were duplicated across different sites or timepoints,  underscoring the need for reliable and accurate ASV methods in healthcare contexts to verify whether an unknown voice belongs to a known enrolled individual.}
Healthcare systems are increasingly relying on automatic speaker verification (ASV) models to ensure secure and accurate identification of patients and healthcare providers, with the aim of preventing fraud, safeguarding patient privacy, and ensuring the accuracy of medical records \citep{upadhyay2022feature, arasteh2022effect}.

Conducting large-scale clinical trials, involving numerous patients, doctors, clinics, and even different countries can pose significant challenges in identifying instances of duplicate participation, which occurs when a single individual joins the same study more than once, either at different sites or time points, leading to skewed results and undermining the validity of study findings \citep{irum2019speaker}.
\citet{shiovitz2013cns} discovered that as much as 7.78\% of patients involved in a clinical study were duplicated across different sites. 
 % \MS{\sout{According to reports from pharmaceutical corporations, there is a duplication rate of 1.5 to 5 percent among the participants monitored during the development process of a single compound \citep{shiovitz2011failure}.}} 
 
 In some cases, individuals participate in multiple clinical trials concurrently in order to earn more money. 
% To make study coordinators think they are genuine, they may provide fallacious answers and pretend to be compliant, resulting in higher placebo rates and compromised data integrity. 
 When a trial enrolls an adequate number of substandard participants, it risks not meeting the primary endpoints and ultimately causing a multimillion-dollar study to fail. \citet{pinho2021improving} examined the financial effect of duplicate participants on the pharmaceutical companies conducting a set of short-term study programs across psychiatric disorders including Schizophrenia, Major Depressive Disorder, and Bipolar Depression. Based on their results, enrolling ineligible subjects in the selected studies results in a loss of around \$29,680,000 for the sponsor pharmaceutical company. In addition, duplicate participation results in higher placebo rates and compromised data integrity. These findings highlight the importance of addressing the duplicate participant problem and underscore the need for reliable and accurate ASV methods in healthcare systems to verify whether an unknown voice belongs to a known enrolled individual \citep{upadhyay2022feature, arasteh2022effect}. % ASV models can be categorized into text-dependent (TD), which involves using the same text for both enrollment and evaluation phases or text-independent (TI), which allows for more flexibility in the enrollment and verification utterances without any constraints \citep{tu2022survey}. When pre-trained on large audio datasets, TI models demonstrate comparable accuracy to that of TD models \citep{tu2022survey}.

% Cognitive impairment has been shown to have a detrimental effect on vocabulary richness, syntactic complexity, and speech fluency \citep{thomas2005automatic, roark2011spoken, guinn2012language, meilan2012acoustic}. This raises concerns about whether the abnormal speech patterns of individuals with cognitive impairment could impact the performance of ASV. However, there is a lack of research on the relationship between cognitive impairment and ASV in the existing literature. This motivates us to address this gap and examine the impact of Alzheimer's disease (AD) severity level on ASV performance. Additional external factors including but not limited to participant profile \citep{si2021exploring}, recording environment, or data collection procedure \citep{woo2006mobile, wan2017speaker} may also affect the ASV performance, but the impact of these factors is understudied in the healthcare field. A comprehensive analysis of these external factors could provide valuable insights into the accuracy and reliability of ASV models and help detect the potential sources of bias that may arise due to differences in the inherent voice characteristics among subgroups \citep{si2021exploring}. Identifying and addressing such sources of errors and bias can enable researchers and practitioners to enhance the fairness and effectiveness of ASV models developed for healthcare applications.

Cognitive impairment has been linked to a decline in vocabulary richness, syntactic complexity, and speech fluency, according to previous research \citep{thomas2005automatic, roark2011spoken, guinn2012language, meilan2012acoustic}. Therefore, it is important to investigate whether the abnormal speech patterns exhibited by individuals with cognitive impairment can affect ASV performance. Despite this concern, there is a paucity of research examining the relationship between cognitive impairment and ASV in the existing literature. This research gap motivated us to address this issue by examining the effect of Alzheimer's disease (AD) severity level on ASV performance. 

Furthermore, external factors such as participants' demographic information \citep{si2021exploring}, recording environment, or data collection procedure \citep{woo2006mobile, wan2017speaker} may also have an impact on ASV performance, but their impact is not well-studied in the healthcare industry. An extensive analysis of these external factors could provide valuable insights into the accuracy and reliability of ASV models and identify potential sources of bias due to differences in inherent voice characteristics among subgroups \citep{si2021exploring}.

% It is important to note that ASV performance may be influenced by various external factors, including participant profile \citep{si2021exploring}, recording environment, and data collection procedure \citep{}. However, research on the impact of these factors on ASV performance in the healthcare field is still limited. As a result, it is crucial to conduct further analysis to better understand the effects of these external factors on ASV performance in clinical trials. Such analysis could provide valuable insights into the factors that may affect the accuracy and reliability of ASV in healthcare settings.

% However, achieving sufficient ASV performance can be problematic when dealing with abnormal speech patterns that are often present in individuals with cognitive impairment \citep{arasteh2022effect}. %To the best of our knowledge, 
% The effect of the severity level of cognitive impairment, such as Alzheimer's diseas (AD) on ASV performance has not been addressed in the prior research although it is investigated whether speech pathology affects ASV performance \citep{arasteh2022effect}.  

The purpose of this study is to investigate the effectiveness of ASV models in identifying duplicate patient participation in large-scale clinical trials, and to explore the factors that influence ASV performance in such settings. To this end, we utilize a longitudinal clinical dataset of English speech recordings obtained through multiple speech tasks from 659 participants with varying levels of AD. We employ the TitaNet model, an end-to-end deep learning text-independent ASV model pre-trained on a large volume of speech recordings of English speakers. ASV models can be classified into two groups: text-dependent (TD) and text-independent (TI). TI ASV models allow for more flexibility in the enrollment and verification phases without constraints on the speech content. When pre-trained on extensive audio datasets, TI models demonstrate a comparable level of accuracy to TD models. We evaluate the performance of TitaNet on our dataset in a zero-shot setting, achieving a 3.1\% equal error rate (EER). In addition, we analyze the impact of various external factors on ASV performance, including participant demographic characteristics (i.e., age, and gender), audio quality criteria (i.e., clinician interference, background noise, participant accent, and participant clarity), as well as AD severity level. This study aims to provide valuable insights into the factors that can affect the performance of ASV models in clinical trial environments, with the goal of improving the accuracy, fairness, and reliability.

Our findings indicate that voice biometrics may present fairness issues, as certain subgroups demonstrate differing speaker verification performances due to their inherent voice characteristics.  In addition, the quality of speech recordings can impact ASV performance, highlighting the importance of monitoring and enhancing data collection and recording settings during clinical trials. 

% By addressing these concerns, the accuracy, reliability, and fairness of ASV models can be improved, thereby increasing their effectiveness in healthcare applications.

\section{Related Work}
Speaker verification technology has been increasingly utilized in various domains, including healthcare systems. Several studies have analyzed the feasibility and effectiveness of speaker verification models in healthcare settings \citep{haovoice, weng2014speaker}. However, the  external factors that can affect the performance of ASV models has not been extensively studied through research in the healthcare field. 

% \MS{Q: I would suggest to order them always in the same way. So far, your order was demographics, audio quality, AD severity level, so you would put Speech Pathology Effect as the last section.}

\paragraph{Race and Gender Effect:} 
% \MS{Q: What is the result of SI2021 study? Which diseases did they cover?}
\citet{si2021exploring} utilized three state-of-the-art ASV models including the Xvector-TDNN \citep{snyder2018x}, ECAPA-TDNN \citep{desplanques2020ecapa}, and DTW \citep{dutta2008dynamic} models to explore demographic effects on speaker verification. For this purpose, they used a subset of the mPower study \citep{bot2016mpower}, a Parkinson's disease mobile dataset, comparing a diverse group of 300 speakers by race and gender. Their results demonstrated that the Latinx subgroup indicates the worst ASV performance among the four major races in the dataset (i.e., White, Black, Latinx, and Asian). Based on their findings,  gender represents minor differences in
ASV performance between male-only and female-only subgroups. We did similar gender-level and accent-level analyses on patients with Alzheimer's disease to detect the potential sources of bias in ASV due to the inherent voice characteristics of distinct genders or English accents. 

\paragraph{Age Effect: }\citet{kelly2011effects} analyzed the effect of long-term ageing on ASV performance. They utilized a conventional GMM-UBM system \citep{irum2019speaker} on a longitudinal voice dataset of a cohort of 13 adult speakers, whose recordings were collected over a time span of 30-40 years. According to their results, short-term aging (less than 5 years) does not have a significant impact on verification performance, compared to normal inter-session variations. However, for longer periods, aging has a negative effect on verification accuracy. Moreover, the researchers found that the rate of verification score decline is more rapid for speakers aged 60 years and above. However, they only evaluated their models on small cohorts and inter-speaker differences across different age groups were not further analyzed, while in the present work, we evaluate the ASV performance across different age groups over 55 years old and incorporate a larger clinical dataset with 659 speakers in total, while we controlled for AD effect. \citet{taylor2020age} also demonstrated that some speech and vocal characteristics (e.g., the spectral center of gravity, spectral skewness, or spectral kurtosis) undergo alterations with aging, and these changes can vary between men and women. These findings suggest that age is an effective factor in speaker's voice characteristics and this underscores the importance of assessing age effect on our ASV model to ensure the fairness of the model across different age groups.

\paragraph{Noise Effect: }\citet{wan2017speaker} applied LibriSpeech corpus \citep{panayotov2015librispeech} of English novel reading speech with varying lengths and tested ASV performance across different types and levels of background noise (e.g., babble, car, office and airplane noise) with a great mismatch between training and testing speech.  Based on their findings, performance varies across different types of noise and the number of errors grow with a decrease in the sound-to-noise-ratio value. However, other metrics of audio quality were not considered in their study and their models were only trained and tested on healthy speech recordings. In the present study, we assess the effect of other audio quality aspects, such as participant clarity, clinician interference, and background noise, on ASV performance using a dataset of speakers with varying severity levels of AD, which was collected in clinical environment.

\paragraph{Speech Pathology Effect: } \citet{arasteh2022effect} have investigated the vulnerability of pathological speech to re-identification in ASV systems. In a large-scale study, they explored the effects of different speech pathologies on ASV using a real-world pathological speech corpus of more than 2,000 test subjects with various speech and voice disorders.  %The authors utilized a deep learning-based TI ASV model and obtained a mean EER of 0.89±0.06\%, with n=2,064 speakers in the training dataset and n=517 in the test dataset. 
Their results indicated that some types of speech pathology, particularly dysphonia, regardless of speech intelligibility, are more vulnerable to a breach of privacy compared to healthy speech. They do not analyze the effect of AD on ASV performance, even though speech and language impairment are prevalent issues in moderate to severe stages of AD that may potentially affect the ASV performance. This motivates us to evaluate ASV performance across varying severity levels of AD.

% As patients progress from moderate to severe disease stages, language impairment becomes a significant concern for most of them.

% The performance of their model is dependent on a sufficient amount of labeled clinical speech segments available for the training phase. In addition, they don't include any types of cognitive impairment which commonly 

% Instead, we utilize a pre-trained model in a zero-shot setting with no fine-tuning or labeled training examples required. 

\section{Methods}
\subsection{Datasets}
The Alzheimer's Disease Clinical Trial (ADCT) dataset comprises speech recordings of English-speaking patients with a clinical diagnosis of mild to moderate AD who participated in a clinical trial. This is a proprietary dataset, which was collected every 12 weeks for a 48-week treatment period. It includes recordings of participants performing a set of self-administered speech tasks, including picture description \citep{goodglass2001bdae, becker1994natural}, phonemic verbal fluency \citep{borkowski1967word}, and semantic verbal fluency \citep{tombaugh1999normative} tasks. 

\subsubsection{Demographic Information}
Demographic data were collected about the participants at the beginning of the study. This data includes the age, and gender of the individuals upon consenting. The data collection study was approved by the ethical committee. 

% The study was performed in accordance with US IND
% regulations (21 Code of Federal Regulations [CFR] 312) \citep{}, US local national laws (as applicable) \citep{}, ICH guidelines for Good Clinical Practice \citep{}, and the most recent guidelines of the Declaration of Helsinki \citep{}.

\subsubsection{Transcription and Quality Assessment}
All the audio recordings were manually transcribed by 49 trained transcriptionists based on the CHAT protocol and annotations \citep{macwhinney2014childes}. The transcriptionists utilized an online tool that granted them access to the recordings and enabled them to transcribe the audio content, segment the files into utterances, and perform quality assessment. In addition, the transcriptions manually rated the quality of the recordings according to different quality criteria. The values range from 0 to 3 for each quality criterion. Value higher than 0 indicates that the audio recording has minor to major issues under that quality criterion. The quality criteria consist of background noise, clinician interference, participant accent, and participant clarity. The background noise criterion indicates whether there is noise in the background from the environment. Clinician interference indicates whether the clinician (or another speaker) interferes with the speech task. The participant accent criterion indicates whether the participant is a native or near-native speaker (values of 0) or has a detectable non-native accent (value higher than 0). Participant clarity indicates whether the participant’s voice is hard to hear or understand.

\subsubsection{Clinical Assessment}
Participants were assessed on the severity level of AD using the Mini-Mental State Examination (MMSE) \citep{henneges2016describing} rating scale. MMSE is a brief cognitive function assessment, which consists of 30 questions that can be completed in less than 10 minutes. The questions are divided into seven categories, with each subscore examining a particular aspect of cognition: Orientation in time (score range 0-5), orientation in place (score range 0-5), registration (score range 0-3), attention and concentration (score range 0-5), recall (score range 0-3), language (score range 0-8), and drawing (score range 0-1). MMSE total score ranges from 0 to 30, with higher scores indicating better cognitive function and lower scores indicating more severe cognitive impairment. In this study, the participants were categorized into four levels of AD severity based on MMSE criteria~\citep{wimo2013geras}: Healthy Control (HC) (MMSE score > 26 points), Mild AD (MMSE score 21-26 points), Moderate AD (MMSE score 15-20 points), and Severe AD (MMSE score < 15 points).

\subsubsection{Dataset Composition}
\label{sec:data-composition}
The ADCT data comprises 7084 audio recordings from 659 speakers with 10.7$\pm$7.0 samples on average per each speaker. The average duration of total audio and speech-only audio are equal to 69.31 and 37.30 seconds, respectively.

\begin{figure}[t]
    \centering
    \includegraphics[width=1\linewidth]{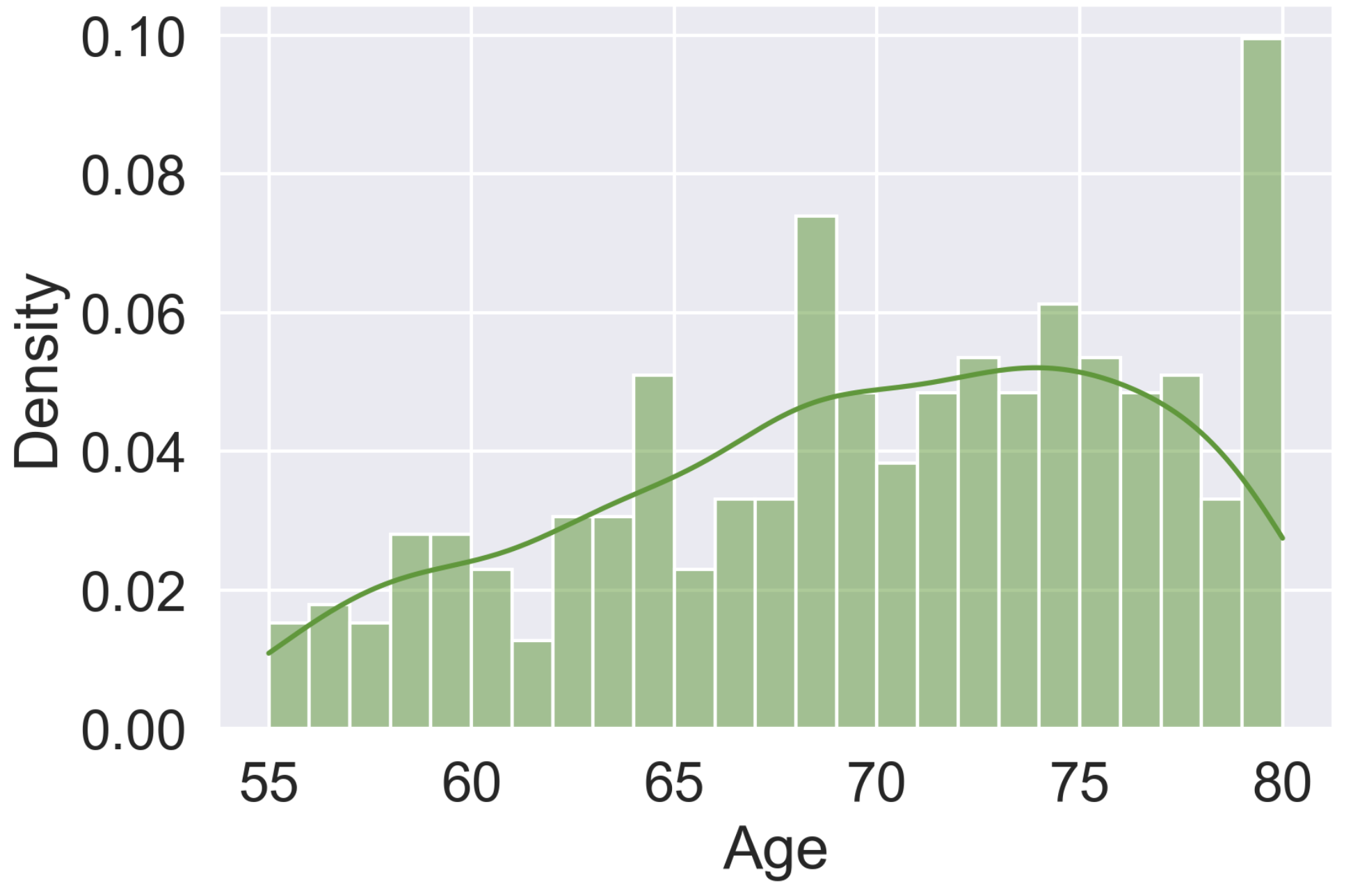}
    \caption{Age distribution of the speakers in ADCT dataset.}
    \label{fig:age-dist}
\end{figure}

In the dataset, 43.4\% of the speakers are male and 56.6\% of the speakers are female. The age range of the subjects spans from 55 to 80 years old. Age distribution of the subjects is represented in Figure \ref{fig:age-dist}, with an average value equal to 69.7$\pm$6.7. 

\begin{figure}[t]
    \centering
    \includegraphics[width=1\linewidth]{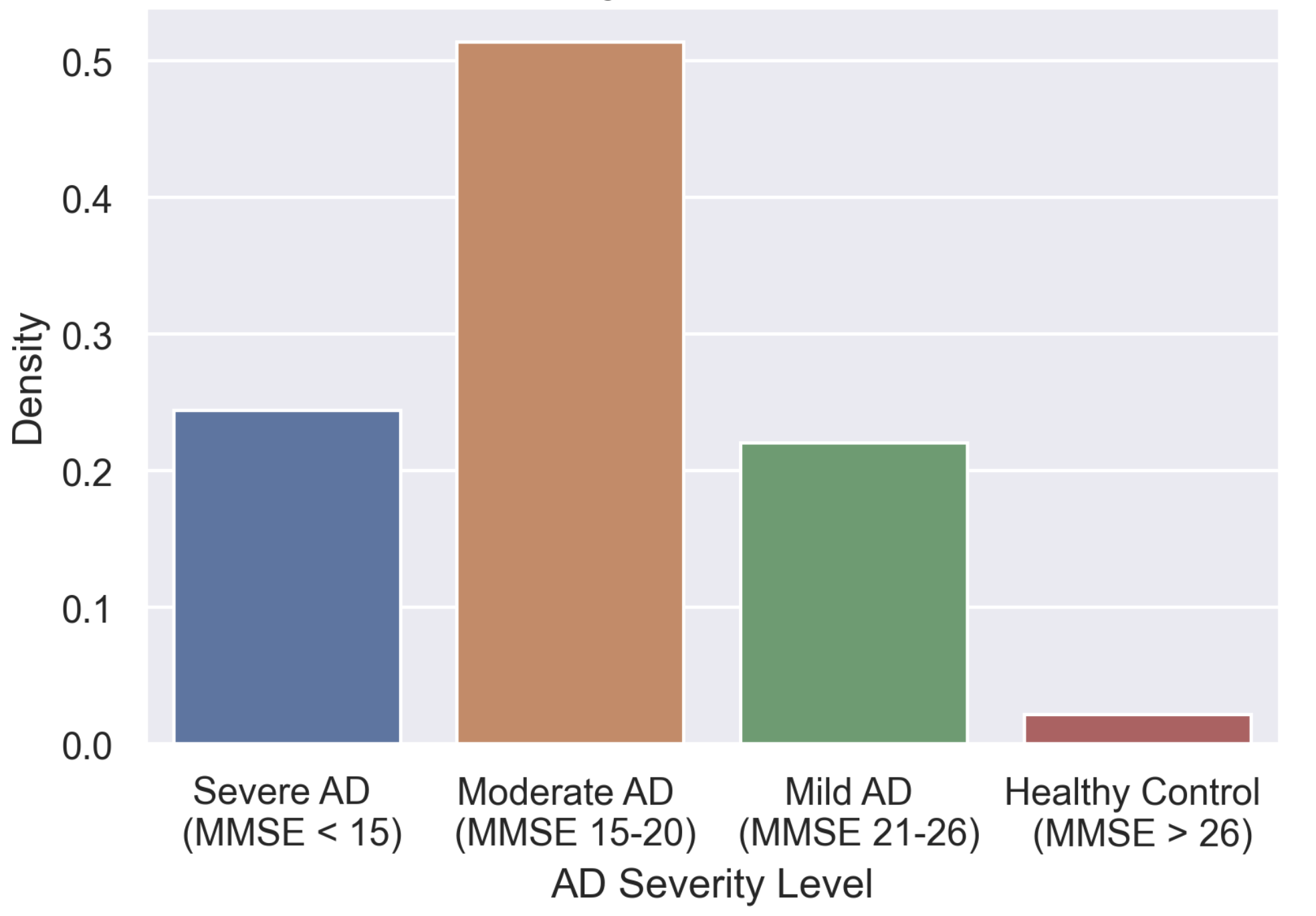}
    \caption{Distribution of AD severity levels in ADCT dataset.}
    \label{fig:mmse-dist}
\end{figure}

Figure \ref{fig:mmse-dist} indicates the distribution of MMSE scores in the ADCT dataset, showing that the majority of the samples consist of mild to severe levels of AD with scores in the range of 15 to 26 points. The average MMSE score is equal to 17.3$\pm$4.4. It should also be noted that the severity level of AD may vary over time for some of the speakers.

% \subsection{Automated Audio Quality Enhancement}
%Experiments are in progress
\subsection{Models}

% In this study, we applied a state-of-the-art end-to-end TI ASV models from the Nvidia NeMo toolkit\footnote{https://github.com/NVIDIA/NeMo} that had been pretrained on extensive collection of English speech data, from various publicly-available resources. The selected models include SpeakerNet \citep{koluguri2020speakernet}, TitaNet \citep{koluguri2022titanet}, and ECAPA-TDNN \citep{dawalatabad2021ecapa} models. We banchmarked their performance on ADCT dataset in a zero-shot setting and picked the best-performing model achieving the lowest EER, which was TitaNet, for the multifactor analysis of ASV performance.

In this study, we utilized the TitaNet model \citep{koluguri2022titanet}, which is a state-of-the-art end-to-end TI ASV model from the Nvidia NeMo toolkit\footnote{https://github.com/NVIDIA/NeMo} that had been pre-trained on an extensive collection of English speech data, from various publicly-available resources. 
% \begin{figure}[t]
%     \centering
%     \includegraphics[width=0.39\textwidth]{figures/TitaNet.png}
%     \caption{TitaNet model architecture \citep{koluguri2022titanet}} 
%     \label{fig:titanet_diagram}
% \end{figure}
The TitaNet model is a neural network model that adopts an encoder-decoder architecture to extract speaker embeddings from speech. The model architecture is inspired by ContextNet \citep{han2020contextnet} model, which comprises 1D depth-wise separable convolutions followed by squeeze and excitation (SE) layers combined with channel attention pooling to convert utterances of varying lengths into a fixed-length embedding. The model contains 25.3M parameters and it is pre-trained on the VoxCeleb1 Dev \citep{nagrani2017voxceleb}, VoxCeleb2 Dev \citep{chung2018voxceleb2}, Fisher \citep{cieri2004fisher}, Switchboard-Cellular1, Switchboard-Cellular2 \citep{godfrey1993switchboard}, and LibriSpeech \citep{panayotov2015librispeech} datasets.

 We applied the model to the ADCT dataset in a zero-shot setting and used it for further analysis of the effect of external factors.
 
\subsection{Evaluation}

To analyze the effect of  external factors on ASV performance, we separately evaluated the performance of the TitaNet model across different subsets of ADCT data including genders, age groups, audio quality levels, and AD severity levels. Initially, we produced embeddings for all audio files within each group. Subsequently, we aggregated these embeddings to create a set of tuples comprising positive and negative pairs.

\textbf{Positive tuples} refer to pairs of embeddings that belong to the same speaker.  The total number of positive tuples in each group is given by $\sum_{i=1}^m {n_i \choose 2}$, which is calculated by summing up all pairs of $n_i$ speech recordings for the i\textsuperscript{th} speaker, where $m$ is the total number of speakers in the same group. 

% \AA{\sout{\textbf{Negative tuples} refer to pairs of embeddings that belong to different speakers within the same group. The total number of negative pairs for a speaker is calculated by multiplying the number of speech recordings for the speaker with the number of speech recordings for all other speakers in the same group. The sum of all these products for all speakers divided by 2 gives $\sum_{i=1}^m {n_i * (N-n_i)/2}$ negative pairs in total, where $N$ is the number of all audio files in the same group. The sum of products is divided by 2 because each pair of embeddings is counted twice for each speaker in the tuple.}}

\textbf{Negative tuples} refer to pairs of embeddings from different speakers within the same group. The total number of negative pairs is calculated as $\sum_{i=1}^m {n_i * (N-n_i)/2}$, where $N$ is the total number of audio files in the group, and $n_i$ is the number of speech recordings for speaker $i$. The sum is divided by 2 to avoid counting each pair twice.

After generating all the positive and negative pair tuples, we proceeded to compute the cosine similarity between the pairs of vector embeddings within each tuple. Subsequently, we adjusted a threshold value $\theta$ for each group through manual tuning until the true positive rate equalled the true negative rate, which enabled us to evaluate the performance of the ASV  model using the equal error rate (EER) metric. If the cosine similarity value exceeded the threshold, we considered the corresponding tuple as belonging to the same speaker. Conversely, if the cosine similarity value was below the threshold, we deemed the two embeddings to represent different speakers.

% There were around $m$ times as many negative tuples as there were positive tuples due to there being more ways of creating pairs of negative tuple pairs than positive pairs.

\section{Results and Discussion}
\begin{table*}[htbp]
\centering
%\scriptsize
%\small
      \begin{tabular}{llllllll}
          \hline
          \multirow{2}{*}{\bfseries Gender} & \bfseries Tuned & \multirow{2}{*}{\bfseries EER(\%)} & \multirow{2}{*}{\bfseries \#Spkrs} & \multirow{2}{*}{\bfseries \#Smpls} & \bfseries Avg \#Smpls & \bfseries Avg Age & \bfseries Avg MMSE\\
          \bfseries & \bfseries Thr. & \bfseries & \bfseries & \bfseries & \bfseries per Spkr & \bfseries & \bfseries Score \\
          \hline \hline
          \textbf{Female} & 0.75 & 5.13  & 170 & 2735 & 16.09$\pm$3.94 & 69.53$\pm$6.72 & 17.33$\pm$4.37 \\ 
          \textbf{Male} & 0.75 & 4.98  & 170 & 2671 & 15.72$\pm$4.02 & 69.41$\pm$6.96 & 17.45$\pm$4.45\\ 
          \textbf{All} & 0.74 & \textbf{3.10} & 659 & 7084 & 10.70$\pm$7.00 & 69.55$\pm$6.75 & 17.32$\pm$4.44\\
          \hline 
      \end{tabular}
  \caption{Benchmarking TitaNet ASV model across different genders on the ADCT English dataset. `\#Spkrs' denotes the number of speakers per each gender. `\#Smpls' denotes the number of audio recordings per each gender. `Tuned Thr.' denotes tuned threshold.}
  \label{tab:gender-performance}
\end{table*}

\begin{table*}[htbp]
\centering
%\scriptsize
%\small
      \begin{tabular}{llllllll}
          \hline
          \multirow{2}{*}{\bfseries Age Group} & \bfseries Tuned & \multirow{2}{*}{\bfseries EER(\%)} & \multirow{2}{*}{\bfseries \#Spkrs} & \multirow{2}{*}{\bfseries \#Smpls} & \bfseries Avg \#Smpls & \bfseries Gender & \bfseries Avg MMSE\\
          \bfseries & \bfseries Thr. & \bfseries & \bfseries & \bfseries & \bfseries per Spkr & \bfseries & \bfseries Score \\
          \hline \hline
          \textbf{Age <= 70} & 0.73 & 3.62  & 197 & 3235 & 16.42$\pm$3.86 & Male+Female & 17.09$\pm$4.72 \\ 
          \textbf{Age > 70} & 0.74 & 4.20  & 195 & 3022 & 15.50$\pm$4.07 & Male+Female & 17.57$\pm$4.11\\ 
          \textbf{All} & 0.74 & \textbf{3.10} & 659 & 7084 & 10.7$\pm$7.0 & Male+Female & 17.32$\pm$4.44\\
          \hline 
      \end{tabular}
  \caption{Benchmarking TitaNet ASV model across different age groups on the ADCT English dataset. `\#Spkrs' denotes the number of speakers per each age group. `\#Smpls' denotes the number of audio recordings per each age group. `Tuned Thr.' denotes tuned threshold.}
  \label{tab:age-performance}
\end{table*}

In order to have a baseline level of ASV performance, we evaluated the TitaNet model on all the speech recordings of the ADCT dataset and obtained a 3.1\% EER. To further analyze the impact of participant demographic characteristics, audio quality and AD severity level, we then recalculated EER and compared the ASV performance across different subgroups.

\begin{figure}[h]
    \centering
    \includegraphics[width=0.39\textwidth]{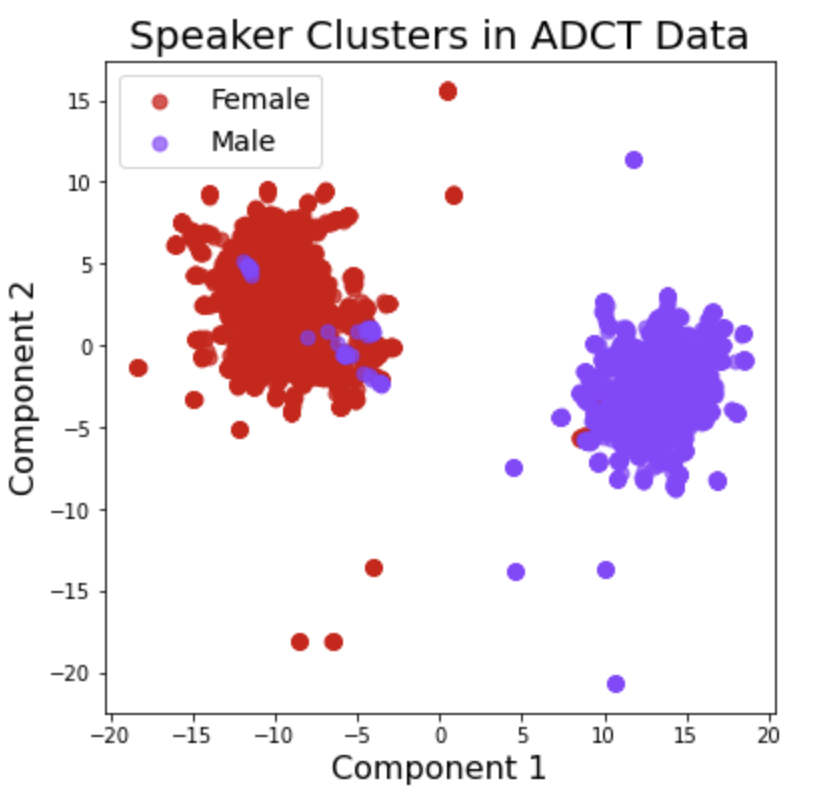}
    \caption{Visualization of speaker clusters using TitaNet embeddings of all audio recordings in ADCT dataset, created using the PaCMAP \citep{JMLR:v22:20-1061} dimensionality reduction method, where each color represents the gender of the speaker.} 
    \label{fig:gender_clusters}
\end{figure}
\subsection{Effect of Participant Demographic Characteristics on ASV Performance}
\subsubsection{Gender Effect}
We first analyzed the effect of speaker genders on the performance of ASV. As shown in Figure \ref{fig:gender_clusters}, two visually distinguishable speaker clusters appeared in the visualization of speech embeddings of all speakers in our dataset. Colouring the data points based on their gender indicates that each cluster is representative of a specific gender. The left cluster mostly comprises female speakers and the majority of the right cluster consists of male speakers. For further analysis of the ASV performance, we separately evaluated the model performance on male and female speech samples within each cluster. To control for the confounding factors, we randomly downsized the size of the female subgroup to the number of speakers in the male subgroup and also, matched the average age and MMSE score between the two subgroups. The results (Table \ref{tab:gender-performance}) show that according to the EER metric, the ASV model performs better on the total dataset comprising diverse genders compared to when it is applied to the male-only or female-only speakers (Table \ref{tab:gender-performance}). The ASV performance for the male subgroup (EER = 4.98\%) is slightly better than that for the female subgroup (5.13\% EER), although the difference is not substantial. This is in line with prior literature \citep{hanilcci2013investigation} demonstrating that male speakers exhibit higher speaker recognition accuracy compared to female speakers regardless of the database and classifier used. The results also align with \citet{si2021exploring} indicating that there is little difference in gender in terms of the performance of ASV models in general.

\subsubsection{Age Effect}
We then evaluated the age effect on ASV performance. For this purpose, we categorized the speakers into two age subgroups with Age <= 70 and Age > 70 according to the speaker ages at the study enrollment date. The threshold was set to 70 because it is equal to the approximate median and mean value of the age distribution of the speakers in the ADCT dataset (Section \ref{sec:data-composition}). To control for the confounding factors, we designed our experiments to ensure that the age-based subgroups had a comparable number of speakers and average MMSE scores and included a combination of male and female speakers. 

Based on the results indicated in Table \ref{tab:age-performance}, EER for participants under 70 is 0.58\% lower than the older age group. These results can be explained by \citet{taylor2020age} revealing that specific attributes of speech and voice characteristics (e.g., fricative spectral moments, semitone standard deviation, etc.) vary according to age. %and that certain changes manifest differently in men and women. 

% Furthermore, the results show that the ASV model demonstrated superior performance, achieving an EER of 3.1\% on the total ADCT dataset, which exhibits greater age diversity. This suggests that the incorporation of a wider range of age groups leads to better ASV performance due to more variations in the inherent voice characteristics of the demographic subgroups.

\subsection{Effect of Audio Quality on ASV Performance}
\begin{table*}[htbp]
\centering
\scriptsize
      \begin{tabular}{lllllllll}
          \hline
          \multirow{2}{*}{\bfseries Audio Quality Criterion} & \bfseries Tuned & \multirow{2}{*}{\bfseries EER(\%)} & \multirow{2}{*}{\bfseries \#Spkrs} & \multirow{2}{*}{\bfseries \#Smpls} & \bfseries Avg \#Smpls & \bfseries Gender & \bfseries Avg Age &\bfseries Avg MMSE\\
          \bfseries & \bfseries Thr. & \bfseries & \bfseries & \bfseries & \bfseries per Spkr & \bfseries & \bfseries & \bfseries Score \\
          \hline \hline
          \textbf{Background Noise - No Issue} & 0.75 & \textbf{2.90}  & 125 & 426 & 3.40$\pm$1.45 & M + F & 69.60$\pm$6.72 & 16.94$\pm$5.83\\ 
          \textbf{Background Noise - Minor to Major Issue} & 0.74 & 3.54 & 125 & 511 & 4.08$\pm$2.08 & M + F & 69.21$\pm$6.45 & 16.78$\pm$5.58\\ 
          \hline 
          \textbf{Participant Clarity - No Issue} & 0.75 & \textbf{2.85} & 112 & 481 & 4.29$\pm$1.70 & M + F & 69.80$\pm$6.38 & 16.81$\pm$5.62\\ 
          \textbf{Participant Clarity - Minor to Major Issue} & 0.74 & 3.41 & 112 & 432 & 3.85$\pm$1.83 & M + F & 69.23$\pm$6.81 & 16.04$\pm$5.54\\ 
          \hline
          \textbf{Clinician Interference - No Issue} & 0.75 & \textbf{2.90} & 103 & 659 & 4.30$\pm$2.08 & M + F & 69.40$\pm$6.77 & 17.65$\pm$5.61\\ 
          \textbf{Clinician Interf. - Minor to Major Issue} & 0.73 & 3.38 & 103 & 399 & 3.87$\pm$1.86 & M + F & 69.43$\pm$6.84 & 14.77$\pm$5.22\\ 
          \hline
          \textbf{Participant Accent - Native} & 0.74 & 2.97 & 188 & 901 & 4.79$\pm$2.82 & M + F & 68.63$\pm$6.89 & 17.22$\pm$5.01\\ 
          \textbf{Participant Accent - Non-Native} & 0.74 & \textbf{2.01} & 188 & 594 & 3.16$\pm$1.54 & M + F & 70.45$\pm$6.32 & 17.19$\pm$4.56\\ 
          \hline 
          \textbf{All} & 0.74 & 3.10 & 659 & 7084 & 10.7$\pm$7.0 & M + F & 69.55$\pm$6.75& 17.32$\pm$4.44\\
          \hline
      \end{tabular}
  \caption{Benchmarking TitaNet speaker verification model across different levels of audio quality on the ADCT dataset. `\#Spkrs' denotes the number of speakers per each quality subgroup. `\#Smpls' denotes the number of audio recordings per each quality subgroup. `Tuned Thr.' denotes tuned threshold. For each quality criterion, `No Issue' indicates samples with a rating = 0 and `Minor to Major Issue' indicates samples with a rating > 0 (maximum is 3). `M' denotes male speakers and `F' denotes female speakers. Bold font denotes the subgroup yielding best ASV performance in each quality criterion.}
  \label{tab:audio-quality-performance}
\end{table*}

We examined how different audio quality factors affect the performance of speaker verification in clinical environments. We divided ADCT samples into two subgroups based on their quality rating for each criterion: `No Issue' for samples with a rating of 0 and `Minor to Major Issue' for those with a rating higher than 0 (maximum is 3). In order to mitigate the influence of possible confounding factors, we structured our experiments in such a way that the quality-based subgroups had a similar number of speakers and average MMSE score, and included both male and female speakers. Table \ref{tab:audio-quality-performance} shows the comparison of EER values between each pair of subgroups per audio quality criterion. Our results indicate that subgroups of audio samples with no background noise, and high participant clarity yielded lower EER than subgroups with varying levels of background noise and poor participant clarity. Therefore, better control of the data collection setting, along with the use of high-clarity audio recordings with minimal background noise, would be recommended in order to improve ASV performance in clinical trials. Our findings are in line with \citet{eskimez2018front}, who demonstrated that incorporating a DNN-based speech enhancement technique as a front-end noise reduction module can enhance the ASV performance when applied to noisy speech data obtained from real customers.

Our results also suggest that clinician interference can negatively impact ASV performance with 0.48\% increase in EER. Therefore, it is recommended that clinicians refrain from interrupting participants during speech tasks in recording sessions to prevent any decline in performance.

We also evaluated the speakers' accents as a quality indicator, and our results show that ASV performs better on non-native speakers (2.01\% EER) than on native speakers (2.97\% EER). This suggests that ASV performance can be better in trials that involve participants who speak with diverse non-native English accents as a way to identify unique speech characteristics for each individual.

% According to our findings regarding the effect of participant clarity and background noise on ASV performance, enhancing the resolution and clarity of speech recordings and removing background noise could potentially enhance the performance of ASV in clinical trials. 

% Given the degradation in ASV performance in existence of clinician interference, clinicians are advised not to interrupt the participants while they are performing speech tasks during the recording sessions.

\begin{figure}[t!]
    \centering
    \includegraphics[width=0.39\textwidth]{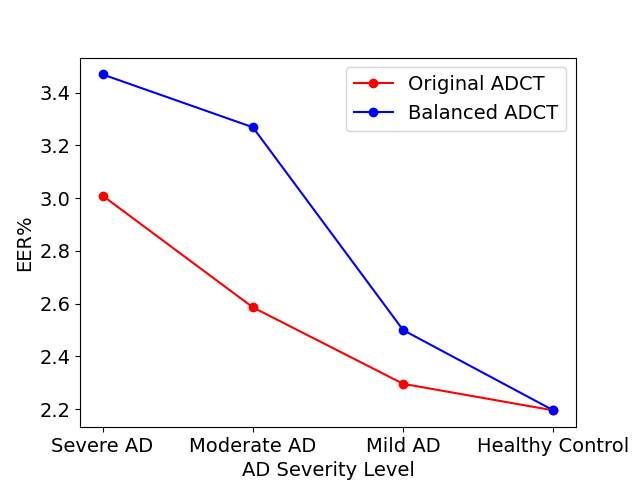}
    \caption{Comparison of the performance of the TitaNet ASV model across different AD severity levels based on EER\% metric. Original ADCT refers to the dataset with the original number of speakers per severity level. Balanced ADCT refers to the dataset with each group downsized to the number of speakers in the smallest group, which is the HC group.}
    \label{fig:severity_effect}
\end{figure}

\subsection{Effect of the Severity Level of Alzheimer's Disease on ASV Performance}

\begin{figure}[h!]
    \centering
    \begin{subfigure}{0.4\textwidth}
    \centering
        \includegraphics[width=5.5cm]{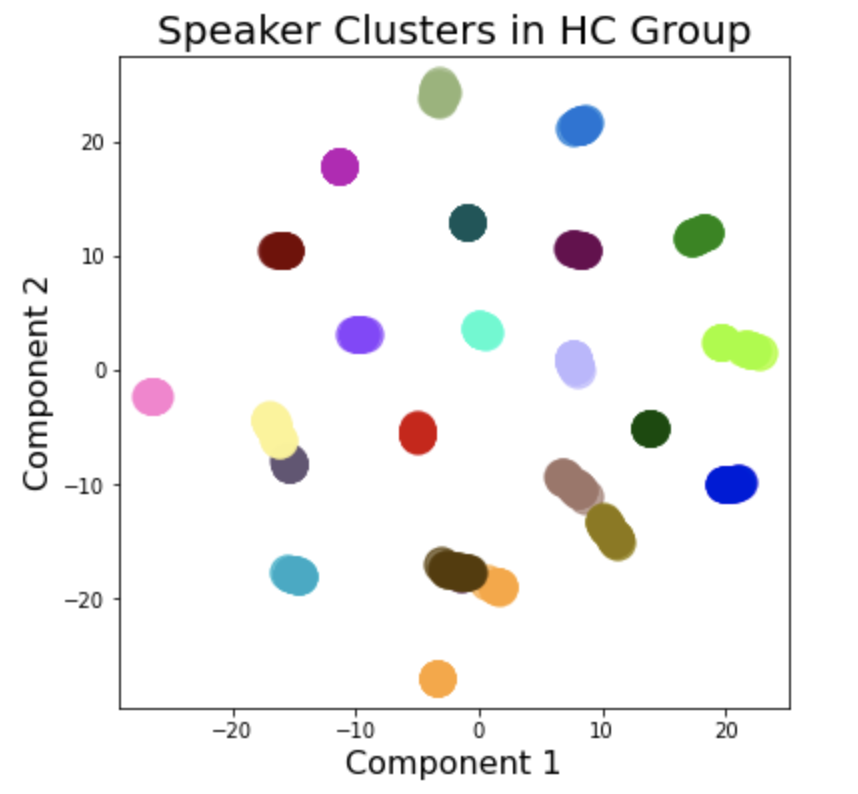}
        \caption{Healthy Control}
        \label{fig:speaker-cluster-hc}
    \end{subfigure}%
    \quad
    \begin{subfigure}{0.4\textwidth}
    \centering
        \includegraphics[width=5.5cm]{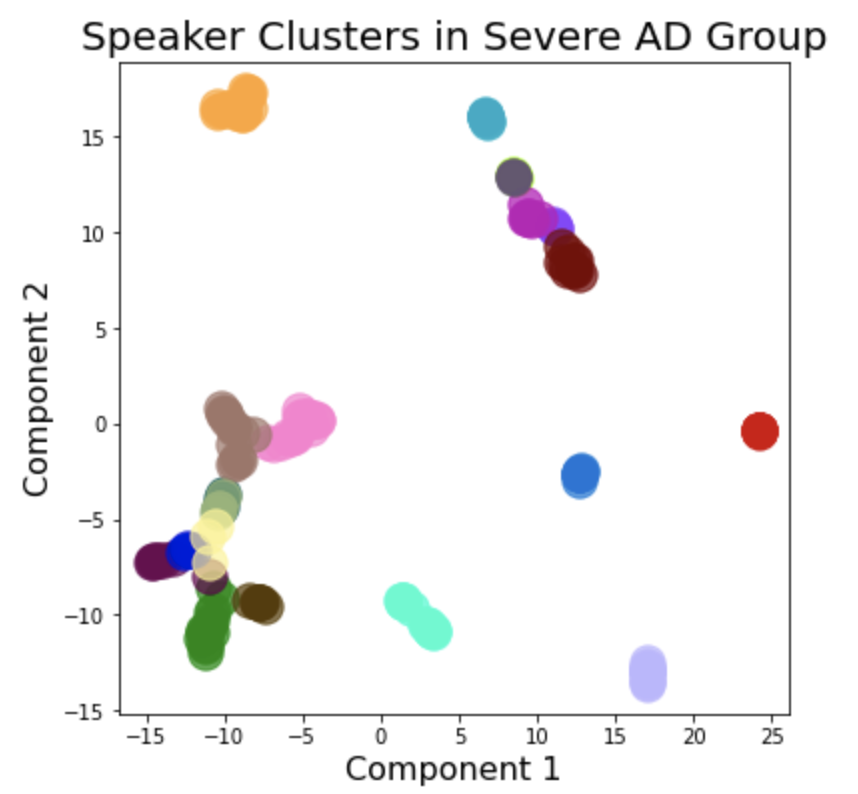}
        \caption{Severe AD}
        \label{fig:speaker-cluster-severe}
    \end{subfigure}
    \caption{Speaker cluster visualizations of HC and Severe AD groups based on TitaNet embeddings for ADCT dataset, created using the PaCMAP \citep{JMLR:v22:20-1061} dimensionality reduction method, with each colour representing a distinct speaker.}
    \label{fig:severity-based-speaker-cluster}
\end{figure}

\begin{figure*} [t]
    \centering
    \begin{subfigure}{0.4\textwidth}
    \centering
        \includegraphics[width=5.2cm]{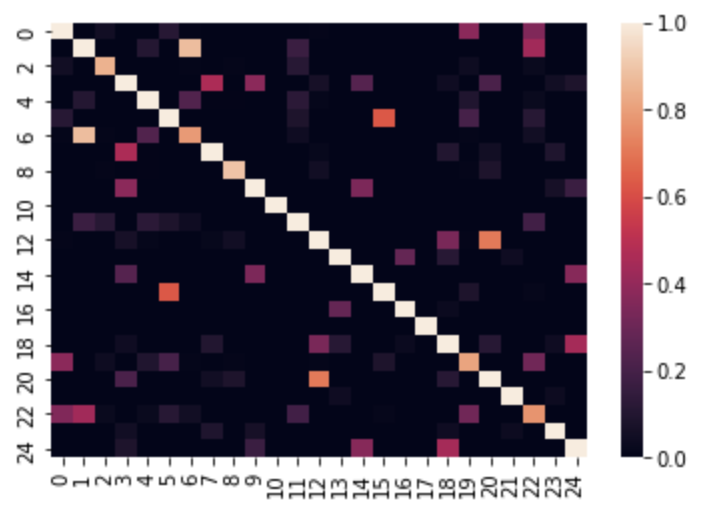}
        \caption{Severe AD}
        \label{fig:heatmap-severe}
    \end{subfigure}%
    \quad
    \begin{subfigure}{0.4\textwidth}
    \centering
        \includegraphics[width=5.2cm]{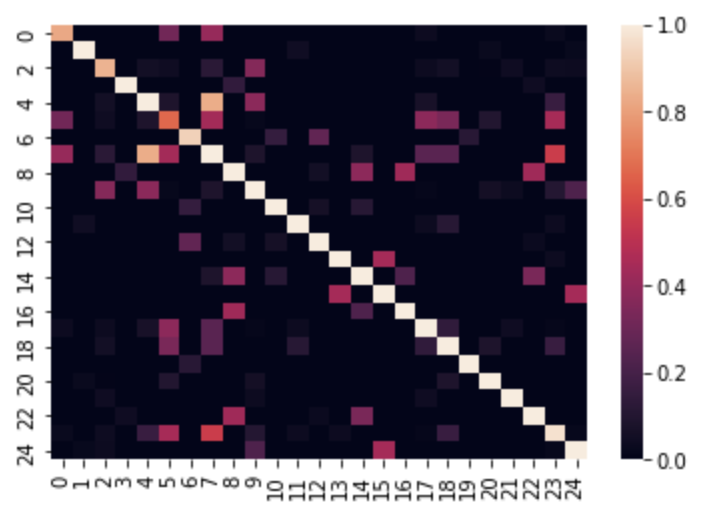}
        \caption{Moderate AD}
        \label{fig:heatmap-moderate}
    \end{subfigure}
    \begin{subfigure}{0.4\textwidth}
    \centering
        \includegraphics[width=5.2cm]{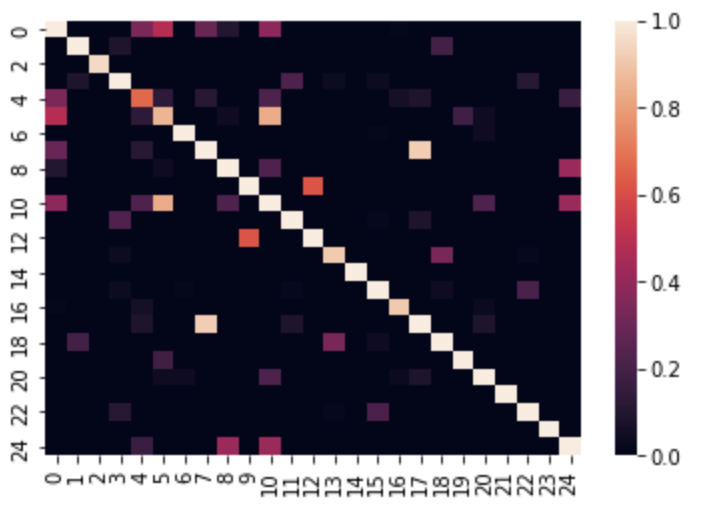}
        \caption{Mild AD}
        \label{fig:heatmap-mild}
    \end{subfigure}%
    \quad
    \begin{subfigure}{0.4\textwidth}
    \centering
        \includegraphics[width=5.2cm]{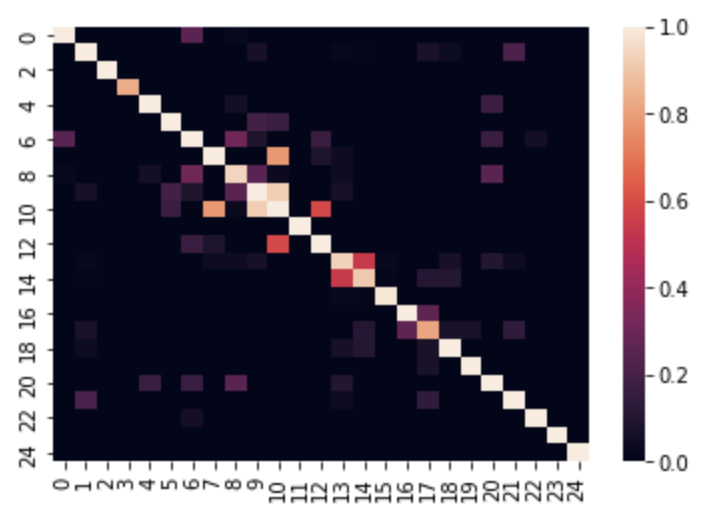}
        \caption{Healthy Control}
        \label{fig:heatmap-hc}
    \end{subfigure}
    \caption{Similarity heatmap visualizations for ADCT dataset using TitaNet embeddings for different AD severity levels each downsized to the number of speakers in the smallest group, which is equal to 25. The lighter colours correspond to a higher level of similarity between the associated speaker tuples.}
    \label{fig:similarity-heatmap}
\end{figure*}

We examined how the AD severity level impacts the performance of the ASV model. For this purpose, we performed a separate recalculation of EER for subgroups of audio samples consisting of Severe AD (Number of speakers = 218), Moderate AD (Number of speakers = 436), Mild AD (Number of speakers = 244), and HC (Number of speakers = 25), while retaining the original number of speakers. To establish a fair comparison, we then balanced the number of speakers across the subgroups by downsizing each to the smallest subgroup size of 25 speakers, which was the size of the HC subgroup. In both scenarios, higher AD severity levels lowered speaker verification performance, by about 1\% to 1.5\% of EER (Figure \ref{fig:severity_effect}). Also, EER was lower within the groups where a higher number of speakers were included. Figure \ref{fig:speaker-cluster-hc} and \ref{fig:speaker-cluster-severe} display the PaCMAP~\citep{JMLR:v22:20-1061} visualization of TitaNet speaker embeddings for the HC and Severe AD groups in the balanced ADCT dataset, %. The figures are plotted by PaCMAP \citep{JMLR:v22:20-1061} dimensionality reduction technique, with 
where each colour representing a unique speaker. In both the HC and Severe AD groups, samples from the same speakers are clustered close to each other, while in the HC group, the samples from different speakers are more distinguishable compared to the Severe AD group. Moreover, there is a higher level of similarity between negative speaker tuples within Moderate and Severe AD subgroups in comparison to Mild AD and HC subgroups, as indicated in the similarity heatmaps for different AD severity levels (Figure \ref{fig:similarity-heatmap}). Overall, our findings suggest that the unique voice characteristics associated with varying levels of AD severity \citep{boschi2017connected} can be  entangled with the identity of the speaker and may introduce a potential source of bias in the ASV models.

\section{Conclusion}
Large-scale clinical trials require accurate verification of participants, as duplicate participation may lead to substandard data quality and significant financial and health risks. Therefore, developing accurate ASV models for verifying participant identity is essential in these settings. External factors such as participant profile or audio quality can cause errors and biases in ASV performance during the trials, but limited research has been conducted in this area. In the present work, we utilize a longitudinal speech dataset of participants with varying levels of AD severity and investigate the impact of external factors, such as different participant demographic characteristics, audio quality criteria, and AD severity levels, on the performance of an end-to-end TI ASV model. Our findings show that variations in ASV performance can be attributed to the inherent voice characteristics of different subgroups (e.g., different ages, genders, accents, or AD severity levels) that are likely to be confused with the identity of the speaker. Hence, it is critical to reassess this technology to mitigate the risk of potential biases toward certain subgroups. Based on our results, poor audio quality with unclear speech, noisy background, and clinician interference also negatively impacts ASV performance. This highlights the importance of quality assurance for the speech recordings during the trials. In future work, we aim to automate the audio quality assessment process by leveraging existing automated methods such as perceptual evaluation of speech quality (PESQ) \citep{rix2001perceptual} or short-time objective intelligibility (STOI) \citep{taal2010short}, which would reduce the human effort required for this task.

\bibliography{anthology,custom}
\bibliographystyle{acl_natbib}

% \appendix

% \section{Example Appendix}
% \label{sec:appendix}

% This is an appendix.

\end{document}